\documentstyle[12pt]{article}

\newfont{\twelvemsb}{msbm10 scaled\magstep1}
\newfont{\eightmsb}{msbm8}
\newfam\msbfam
\textfont\msbfam=\twelvemsb
\scriptfont\msbfam=\eightmsb
\catcode`\@=11
\def\Bbb{\ifmmode\let\next\Bbb@\else
  \def\next{\errmessage{Use \string\Bbb\space only in math mode}}\fi\next}
\def\Bbb@#1{{\fam\msbfam{{#1}}}}

\newtheorem{theorem}{Theorem}

\newcommand{\be}{\begin{equation}}
\newcommand{\ee}{\end{equation}}
\newcommand{\ba}{\begin{eqnarray}}
\newcommand{\ea}{\end{eqnarray}}

\newcommand{\la}{\lambda}

\newcommand{\NP}[1]{Nucl.\ Phys.\ {\bf #1}}
\newcommand{\PL}[1]{Phys.\ Lett.\ {\bf #1}}

\newcommand{\CMP}[1]{Comm.\ Math.\ Phys.\ {\bf #1}}

\newcommand{\MPL}[1]{Mod.\ Phys.\ Lett.\ {\bf #1}}

\newcommand{\TMP}[1]{Teor.\ Math.\ Phys.\ {\bf #1}}
\newcommand{\LMP}[1]{Lett.\ Math.\ Phys.\ {\bf #1}}

\begin{document}
\sloppy
\renewcommand{\thefootnote}{\fnsymbol{footnote}}

\newpage
\setcounter{page}{1}

\vspace{0.7cm}
\begin{flushright}
DCPT-01/57\\
51/2001/EP\\
HWM 01-20\\
EMPG-01-08
\end{flushright}
\vspace*{1cm}
\begin{center}
{\bf From the braided to the usual Yang-Baxter relation}\\
\vspace{1.8cm}
{\large D.\ Fioravanti $^a$ and M.\ Rossi $^b$ \footnote{E-mail:
Davide.Fioravanti@durham.ac.uk, M.Rossi@ma.hw.ac.uk}}\\ \vspace{.5cm}
$^a${\em Department of Mathematical Sciences, University of Durham, United Kingdom} \\
and \\
{\em S.I.S.S.A. and Istituto Nazionale di Fisica Nucleare, Trieste, Italy} \\
\vspace{.3cm}
$^b${\em Department of Mathematics, Heriot-Watt University, Edinburgh,
United Kingdom} \\
\end{center}
\vspace{1cm}

\renewcommand{\thefootnote}{\arabic{footnote}}
\setcounter{footnote}{0}

\begin{abstract}
{\noindent Quantum} monodromy matrices coming from a theory of two
coupled (m)KdV equations are modified  in order to satisfy the usual
Yang-Baxter relation. As a consequence, a general connection  between braided
and {\it unbraided} (usual) Yang-Baxter algebras is derived and also analysed.
 \end{abstract}

\vspace{1cm}
{\noindent PACS}: 11.30-j; 02.40.-k; 03.50.-z

{\noindent {\it Keywords}}: (braided) Yang-Baxter relation; Integrability; Conserved charges.

\newpage

\section{Preliminaries}
The Yang-Baxter relation for the monodromy matrix of a quantum system (\cite{STF,KBI}
and references within) characterizes the integrability of the system. Indeed,
this relation encodes a quantum group {\it symmetry} \cite{Dr1,FRT}, {\it
i.e.} an infinite dimensional deformed Lie algebra including the abelian
conserved quantities -- which enables the system to be {\it \`a la} Liouville
integrable -- as Cartan subalgebra. In some interesting cases this very
rich structure seems to be missing because of non-ultralocal commutation
relations between the fundamental variables (or fields) of the theory.
Nevertheless, a suitable modification of the Yang-Baxter relation was
discovered to hold and to still ensure the Liouville integrability: the
so-called braided Yang-Baxter relation \cite{H,K,FR}. However, this relation
is substantially different from the usual Yang-Baxter relation and without
obvious links to it.

Inside the huge variety of integrable non-ultralocal theories the
prototype is the very interesting case of the quantum (modified) KdV ({\bf
(m)KdV}) system, which gives \cite{G,BLZ,FS} an alternative description of the
minimal Conformal Field Theories ({\bf CFT's}) \cite{BPZ}. The associated
lattice monodromy matrix verifies the braided version of the Yang-Baxter
relation \cite{K} and the transfer matrix can be diagonalized by means of a
generalization of the Algebraic Bethe Ansatz method \cite{FR}. If a left
theory of this kind is properly coupled to a right theory, the resulting
monodromy matrix still satisfies the braided Yang-Baxter relation and was
conjectured in \cite{FR} to give on a cylinder an alternative description of
minimal CFT's perturbed by the $\Phi_{1,3}$ primary operator
\cite{AZ}. On the plane these theories are also described as suitable
restrictions of the sine-Gordon theory \cite {LCL}, which, on the contrary,
exhibits a Yang-Baxter relation \cite {STF}.

In this letter we will address the very natural question of the connection
between  the braided and the usual Yang-Baxter relation. We will use as
examples the cases described in the previous paragraph and involving the
lattice (m)KdV theory as main ingredient. Actually, in what follows we will
show that our treatment is completely general. We will
conclude the letter proposing a new {\it factorised} monodromy matrix
verifying the Yang-Baxter relation.

\medskip

In a previous paper \cite{FR} we considered two copies of the modified KdV
equation,
\be
\partial _{\tau}v=\frac {3}{2}v^2v^\prime +\frac {1}{4}v^{'''}
\quad , \quad   \partial _{\bar\tau}\bar v=\frac {3}{2}\bar v^2\bar v^\prime
+\frac {1}{4}\bar v^{'''} \, , \label {mkdv}
\ee
where $v\equiv-\varphi^\prime$ and $\bar v \equiv-\bar \varphi ^\prime$ are
the spatial derivatives of quasi-periodic Darboux fields defined on
the interval $y$, $\bar y \in[\, 0\, ,\, R\, ]$. The quantizations of the
Darboux fields are the Feigin-Fuks bosons \cite{FF} which satisfy the
commutation relations   \begin{equation}
[\phi(y) \, , \, \phi  (y^\prime)]=-{\frac {i\pi \beta ^2}{2}}s\left (\frac{y-y^\prime}{R}\right)
\quad , \quad   [\bar \phi (\bar y) \, , \, \bar \phi
(\bar y^\prime )]={\frac  {i\pi \beta ^2}{2}}s \left (\frac {\bar y-\bar
y^\prime}{R}\right) \, ,
\label{phi}
\ee
where $\beta ^2>0$ and $s(z)$ is the quasi-periodic extension of the sign
function:
\be
s(z)=2n+1 \quad , \quad n<z<n+1 \quad ;  \quad s(n)=2n \quad , \quad n
\in {\Bbb Z} \, .
\ee
In terms of a discretization of the Feigin-Fuks
bosons,  $\phi_m\equiv \phi\left (m\frac{R}{2N}\right)$,  $\bar \phi_m\equiv
\bar \phi \left (m\frac {R}{2N}\right )$, we defined the $N$-periodic operators
$V_m^{\pm}$ living on a N-site lattice with spacing $\Delta $ and length
$R=N\Delta$:
\ba
V^-_m\equiv \frac {1}{2}\left [ (\phi _{2m-1}-\phi_{2m+1})+(\phi _{2m-2}-\phi  _{2m})
-(\bar \phi _{2m-1}-\bar \phi_{2m+1})+(\bar  \phi _{2m-2}-\bar \phi _{2m})\right]
\label{Ccorr1} \\
V^+_m\equiv \frac {1}{2}\left [(\bar \phi _{2m-1}-\bar \phi _{2m+1})+ (\bar
\phi_{2m-2}-\bar \phi_{2m})-(\phi _{2m-1}-\phi  _{2m+1})+(\phi_{2m-2}-\phi _{2m}) \right] .
\label{Ccorr2}
\ea
These operators are the quantum counterparts of the discretization of the
mKdV variables $v$ and $\bar v$ (\ref {mkdv}).  The exponential operators
$W_m^\pm\equiv e^{iV_m^{\pm}}$ satisfy, as a consequence of (\ref {phi}), the
following exchange relations, first introduced in \cite {FV} (for $m=N$ the
symbols $W^{\pm}_{N+1}$ should be read as $W^{\pm}_1$ respectively):
\begin{eqnarray}
&&W^{\pm}_{m+1}W^{\pm}_m=q^{\pm \frac
{1}{2}}W^{\pm}_mW^{\pm}_{m+1} \, , \, \, W^{\pm}_{m+1}W^{\mp}_m=q^{\mp \frac
{1}{2}}W^{\mp}_mW^{\pm}_{m+1} \, , \, \,  W^+_mW^-_m=qW^-_mW^+_m\, ; \, \,
\nonumber  \\
&& [W_m^{\sharp},W_n^{\sharp^\prime}]=0  \quad {\mbox {if}} \,
\, \, \, \, 2\leq |m-n|\leq N-2    \quad (\,\sharp, \sharp^\prime =\pm \, ) \,.
\label{Wrel}
\end{eqnarray}
By using $W_m^\pm$ we constructed the {\it left} and {\it right} conformal
monodromy matrices
\ba  M(\lambda)&=&L_{N}(\lambda )\ldots L_{1}(\lambda ) \,
, \label {leftmon} \\  \bar M(\lambda)&=&\bar L_{N}(\lambda ^{-1})\ldots \bar
L_{1}(\lambda ^{-1}) \, ,
\label {rightmon}
\ea
together with the off-critical {\it right-left} and {\it left-right}
monodromy matrices, depending on a {\it perturbation} parameter $\mu$
($N\in 4 {\Bbb N}$)
\ba
{\bf M}(\lambda) &\equiv &  \bar L_{N} (\mu^{\frac {1}{2}}{\la}^{-1}) \bar L_{N-1}
(\mu^{\frac {1}{2}}{\la}^{-1})L_{N-2}\left (\mu ^{\frac  {1}{2}}{\lambda}
\right)L_{N-3}\left (\mu ^{\frac  {1}{2}}{\lambda} \right) \ldots \nonumber \\
&&\ldots   \bar L_{4}( \mu^{\frac {1}{2}}{\la}^{-1}) \bar L_{3} (\mu^{\frac
{1}{2}}{\la}^{-1})  L_{2}(  \mu ^{\frac {1}{2}}{\lambda}) L_{1} (\mu ^{\frac
{1}{2}}{\lambda} ) \, , \label {Rightmon}  \\  {\bf M} ^{\prime}(\lambda)
&\equiv &  L_{N}\left (\mu^{\frac {1}{2}}\la \right) L_{N-1}\left  (\mu^{\frac
{1}{2}}\la \right)\bar L_{N-2}({\mu ^{\frac  {1}{2}}} {\la}^{-1})\bar
L_{N-3}({\mu ^{\frac  {1}{2}}}{\la}^{-1}) \ldots \nonumber \\   && \ldots
L_{4}(\mu^{\frac {1}{2}}\la ) L_{3} (\mu^{\frac {1}{2}}\la )  \bar L_{2}(
{\mu ^{\frac {1}{2}}}{\lambda}^{-1}) \bar L_{1} ( {\mu ^{\frac {1}{2}}}{\lambda}^{-1}) \, .
\label {Leftmon}
\ea
In these formul{\ae} left Lax operators $L_m(\la)$ and right Lax operators
$\bar L_m(\la)$ are
\begin{equation}
L_{m}(\lambda) \equiv \left (
\begin{array}{cc} (W_m^-)^{-1} & \Delta \lambda  W_m^+ \\ \Delta \lambda
(W_m^+)^{-1} & W_m^- \\ \end{array} \right )  \, , \quad \bar L_{m}(\lambda)
\equiv \left ( \begin{array}{cc} (W_m^+)^{-1} &  \Delta \lambda W_m^- \\
\Delta \lambda (W_m^-)^{-1} & W_m^+ \\  \end{array} \right ) \, .
\label{lax}
\end{equation}
Monodromy matrix (\ref {leftmon}) and quantum Lax
operator $L_m(\la)$ were first introduced  in \cite{K}, starting from $V_m$,
$\bar V_m$ and not from $\phi_n$, $\bar \phi_n$. Quantum Lax
operators $L_m(\la)$ and $\bar L_m(\la)$ are
{\it non-ultralocal}, {\it i.e.} their operator entries on nearest neighbouring
sites do not commute. As a consequence, monodromy matrices (\ref {leftmon}-\ref
{Leftmon}) satisfy {\it braided} Yang-Baxter relations, first introduced in
complete generality in \cite{H,K}. For instance we have that $M$ (\ref
{leftmon}) satisfies ($a$ and $b$ are two auxiliary spaces)  \be  R_{ab}\left
(\frac {\la}{\la ^\prime}\right )M_a(\la)Z_{ab}^{-1}M_b(\la ^\prime)= M_b(\la
^\prime)Z_{ba}^{-1}M_a(\la)R_{ab}\left (\frac {\la}{\la ^\prime}\right ) \, .
\label{zyb1}   \ee  Matrix ${\bf M} ^{\prime}$ (\ref {Leftmon}) satisfies the
same relation, while  matrices $\bar M$ (\ref {rightmon}) and ${\bf M}$ (\ref
{Rightmon}) satisfy   (\ref {zyb1}) with $Z_{ab}$ replaced by $Z^{-1}_{ab}$.
The usual numerical $R$-matrix is given by   \be
R_{ab}(\xi )=\left(
\begin{array}{cccc} 1 & 0 & 0 & 0 \\
0 & \frac {\xi^{-1}-\xi }{q^{-1}\xi^{-1}-q\xi} & \frac{q^{-1}-q}{q^{-1}\xi^{-1}-q\xi} & 0 \\
0 &\frac{q^{-1}-q}{q^{-1}\xi^{-1}-q\xi }  & \frac{\xi^{-1}-\xi}{q^{-1}\xi^{-1}-q\xi } & 0 \\
0 & 0 & 0 & 1  \end{array}
\right)  \quad ,  \quad q=e^{-i\pi \beta ^2} \,  ,
\label {Rmat}
\ee
and the numerical matrix $Z_{ab}={\mbox{diag}}(q^{-\frac{1}{2}}, q^{\frac{1}{2}},  q^{\frac{1}{2}},
q^{-\frac{1}{2}})$. In the paper \cite {FR} we also showed that
generators $W_m^\pm$ admit a realization in terms of {\it ultralocal}
generators (canonical variables) $x_m$, $p_m$, $1\leq m \leq N$, forming a position-momentum
Heisenberg algebra
\be
[x_m \, , \, x_n]=0\, , \quad
[ p_m \, ,\, p_n ]=0\, , \quad
[ x_m \, ,\,  p_n]= \frac {i\pi \beta ^2}{2}\delta _{m,n} \quad .
\label{heis}
\ee
Indeed, the operators
\be
W_m^{\pm}=e^{i[\pm (x_{m+1}-x_m)-p_m]}
\label {Wreal}
\ee
satisfy exchange relations (\ref {Wrel}). Of course, the symbol $x_{N+1}$ appearing in
(\ref {Wreal}) for $m=N$ should be read as $x_1$.

\medskip

Using this realization we will establish a connection between the monodromy
matrices (\ref {leftmon}-\ref {Leftmon}) satisfying braided Yang-Baxter
relations and some monodromy matrices satisfying the Yang-Baxter relation.

\section{From the braided to the usual Yang-Baxter relation: conformal case}
\setcounter{equation}{0}
Let us rewrite the left Lax operator $L_{m}(\lambda)$ (\ref {lax}) using the
realization  (\ref {Wreal}) for $W_m^\pm$. We have that
\be
L_{m}(\lambda) = \left ( \begin{array}{cc} e^{-i(x_m-x_{m+1}-p_m)} & \Delta \lambda  e^{-i(x_m-x_{m+1}+p_m)}
\\ \Delta \lambda  e^{i(x_m-x_{m+1}+p_m)} &  e^{i(x_m-x_{m+1}-p_m)}\\ \end{array} \right )= D_{m+1}U_m(\la) \, ,
\label {decomp}
\ee
where we have defined:
\be
D_m\equiv \left ( \begin{array}{cc} e^{ix_m} & 0
\\ 0 &  e^{-ix_m}\\ \end{array} \right ) \quad , \quad U_m(\la)\equiv \left ( \begin{array}{cc} e^{-i(x_m-p_m)} & \Delta \lambda  e^{-i(x_m+p_m)}
\\ \Delta \lambda  e^{i(x_m+p_m)} &  e^{i(x_m-p_m)}\\ \end{array} \right )
\, . \label{DUdef}
\ee
Of course, $D_{N+1}=D_1$ in (\ref {decomp}) and the matrices $D_m$ and
$U_m(\la)$ depend only on  ultralocal site variables. Therefore, we interpret
formul{\ae} (\ref {decomp}) as the decomposition of the Lax operator $L_m(\la)$ in its ultralocal
components on the lattice sites.

Decomposition (\ref {decomp}) implies the following form for the left monodromy matrix
(\ref{leftmon}):
\ba
M(\la)&=&L_N(\la)L_{N-1}(\la)...L_2(\la)L_1(\la)=[D_1U_N(\la)]\, [D_NU_{N-1}(\la)]...[D_3U_2(\la)]\, [D_2U_1(\la)] \nonumber \\
&=&D_1\, [U_N(\la)D_N]\, [U_{N-1}(\la)D_{N-1}]...[U_2(\la)D_2]\, [U_1(\la)D_1] \,  D_1^{-1} \, .
\ea
This means that the matrix
\be
\tilde M(\la)\equiv D_1^{-1} M(\la) D_1 \label{Mtil}
\ee
can be written as
\be
\tilde M(\la)=\tilde L_N(\la)\tilde L_{N-1}(\la)...\tilde L_2(\la)
\tilde L_1(\la) \label {tildeM} \, ,
\ee
in terms of the site {\it ultralocal} operators
\be
\tilde L_m(\la)\equiv U_m(\la) D_m  \, .
\label{Ltil1}
\ee
By using (\ref {DUdef}) in the definition (\ref {Ltil1}), we obtain the
following expression for $\tilde L_m(\la)$:  \be
\tilde L_m(\la)= q^{-1/4} \left ( \begin{array}{cc} e^{ip_m} & \Delta \lambda  e^{-i(2x_m+p_m)}
\\ \Delta \lambda  e^{i(2x_m+p_m)} &  e^{-ip_m}\\ \end{array} \right ) \, .
\label {Lreal}
\ee
From this expression and from commutation relations (\ref {heis}) it eventually follows
that the ultralocal operators $\tilde L_m(\la)$ fulfill the Yang-Baxter
relation
\be
R_{ab}\left (\frac {\lambda}{\lambda ^\prime}\right)\tilde L_{am}(\lambda)
\tilde L_{bm}(\lambda ^\prime)=
\tilde L_{bm}(\lambda ^\prime)\tilde L_{am}(\lambda)R_{ab}\left(\frac
{\lambda}{\lambda ^\prime} \right) \, , \label{yb}
\ee
where the matrix $R_{ab}$ is given by (\ref {Rmat}).
As a consequence of the ultralocality of $\tilde L_m$, we obtain as usual
that the matrix $\tilde M$, in its turn, satisfies the Yang-Baxter relation as
well:  \be  R_{ab}\left (\frac{\lambda}{\lambda ^\prime}\right)\tilde
M_a(\lambda)  \tilde M_b(\lambda ^\prime)=
\tilde M_b(\lambda ^\prime)\tilde M_a(\lambda)R_{ab}\left(\frac
{\lambda}{\lambda ^\prime} \right) \, .
\label{Myb}
\ee
Therefore, the transfer matrix $\tilde \tau (\la)\equiv {\mbox {Tr}} \tilde M
(\la)$  commutes with itself for different values of the spectral parameter
\be
[\tilde \tau (\la) , \tilde \tau (\la ^\prime) ]=0 \, ,
\ee
and, as a consequence, generates infinite conserved charges in involution.

\medskip

{\bf Observation}: The aforementioned results of this section have been
obtained using the realization  (\ref {Wreal}). However, the
transformation of an algebra defined by the braided Yang-Baxter relation into
another algebra defined by the usual Yang-Baxter relation is more general, as
follows from the following theorem.
\begin{theorem}
Let $M$ be a matrix satisfying the
braided Yang-Baxter relation (\ref {zyb1}). If there exists an invertible
matrix $D$ satisfying the conditions
\be
M_a(\la)=D_b M_a(\la)
D_b^{-1}Z_{ab}  \quad , \quad [D_a D_b, R_{ab}(\xi)]=0 \quad , \quad
[D_a,D_b]=0 \quad  \label {condi}  \ee  and if $R_{ab}$ and $Z_{ab}$ entering
(\ref {zyb1}) satisfy  \be
Z_{ab}R_{ab}(\xi)=R_{ab}(\xi)Z_{ba} \, ,
\label{condi2}
\ee
then $\tilde M(\la)=D^{-1} M(\la) D$ satisfies the Yang-Baxter relation (\ref {Myb}).
\end{theorem}

The proof comes from direct calculations. In the algebraic context the
meaning of the theorem is that the algebra generated by the entries of a
matrix $M$ satisfying the braided Yang-Baxter relation (\ref{zyb1}), with
$R_{ab}$ and $Z_{ab}$ obeying (\ref{condi2}), is isomorphic to the algebra
generated by the entries of the {\it unbraided} matrix $\tilde M(\la)=D^{-1}
M(\la)D$ satisfying the Yang-Baxter relation (\ref{Myb}), if a matrix (of
formal elements) $D$ obeying conditions (\ref {condi}) exists: an {\it
unbraiding} isomorphism. In the (m)KdV theory (\ref {condi2}) is evidently true
and in the realization (\ref {Wreal}) conditions (\ref {condi}) are satisfied
by the matrix $D_1$, defined in (\ref {DUdef}). A still open problem concerns
the possibility of choosing $D$ in such a way that the above isomorphism
extends to the respective Hopf algebras.
Let us finally remark that the most general definition of braided
Yang-Baxter algebras \cite {H} uses two numerical matrices, $Z_{ab}$ and
$\tilde Z_{ab}$: the defining relation is $R_{ab}\tilde
Z_{ba}^{-1}M_aZ_{ab}^{-1}M_b=\tilde Z_{ab}^{-1}M_bZ_{ba}^{-1}M_aR_{ab}$.
However, the aforementioned isomorphism into a Yang-Baxter algebra exists when
condition (\ref {condi2}) holds for $R_{ab}$ and $\tilde Z_{ab}$ and
this reduces the above general defining relation to (\ref {zyb1}).
\medskip

Let us now study in detail the invertible map from the left conformal monodromy
matrix $M$ into the unbraided monodromy matrix $\tilde M$.  From
relation (\ref {Mtil}) and from the definition of $D_1$ (\ref {DUdef})  it
follows that the relations between the matrix elements of $M$ and $\tilde M$,
\begin{equation}  M(\lambda )\equiv  \left ( \begin{array}{cc}  A(\lambda )&
B(\lambda)  \\ C(\lambda)  & D(\lambda) \\ \end{array} \right ) \quad , \quad
 \tilde M(\lambda )\equiv  \left ( \begin{array}{cc}  \tilde A(\lambda )&
\tilde B(\lambda)  \\ \tilde C(\lambda)  & \tilde D(\lambda) \\ \end{array}
\right ) \, ,  \ee  are the following:  \ba  A(\la)=e^{ix_1}\tilde
A(\la)e^{-ix_1} \quad &,& \quad B(\la)=e^{ix_1}\tilde B(\la) e^{ix_1} \, ,
\nonumber \\  C(\la)=e^{-ix_1}\tilde C(\la) e^{-ix_1} \quad &,& \quad
D(\la)=e^{-ix_1}\tilde D(\la)e^{ix_1} \, . \label{ABCD}  \ea  In order to
simplify relations (\ref {ABCD}) we rewrite the matrix $\tilde M$ as follows:
\ba  \tilde M(\la)&=&\tilde L_N(\la)\tilde L_{N-1}(\la)...\tilde L_2(\la)
\tilde L_1(\la) =\nonumber \\  &=& \left [\tilde L_N(\la)\tilde
L_{N-1}(\la)...\tilde L_2(\la) \right ]    \tilde L_1(\la)\equiv \tilde
M^\prime (\la) \tilde L_1(\la) \, . \label{scorp}  \ea  Hence, from (\ref
{ABCD}, \ref {Lreal}, \ref {scorp}) we obtain, with obvious notations:  \be
A(\la)=e^{ix_1}\left [ \tilde A^\prime (\la)e^{ip_1}+\tilde B^\prime(\la)\Delta \la e^{i(2x_1+p_1)} \right] e^{-ix_1}q^{-1/4} \, . \label{rel1}
\ee
The factor $e^{-ix_1}$ commutes with $\tilde A^\prime (\la)$ and $\tilde B^\prime (\la)$
because these depend only on the ultralocal variables of the sites $2, \ldots , N$. The exchange
with the other factors in the square bracket is regulated by (\ref {heis}) and produces
\be
A(\la)=q^{1/2}\tilde A(\la) \, . \label {A}
\ee
The same procedure applied to the other elements of $M$ and $\tilde M$ gives
\ba
&D(\la)=q^{1/2}\tilde D(\la)\, , & \label {D}  \\
&B(\la)=q^{1/2}e^{2ix_1}\tilde B(\la) \quad , \quad
C(\la)=q^{1/2}e^{-2ix_1}\tilde C(\la)\, .&  \label {BC}
\ea
Let us finally remark that from (\ref{A}, \ref{D}) it follows that
\be
\tilde \tau (\la)=\tilde A(\la)+\tilde
D(\la)=q^{-1/2}[A(\la)+D(\la)] = q^{-1/2}\tau (\la) \, ,  \ee
{\it i.e.} the transfer matrix $\tilde \tau (\la)={\mbox {Tr}} \tilde M (\la)$ is {\it proportional} to the transfer
matrix $\tau (\la)\equiv{\mbox {Tr}} M (\la)$. Hence $\tilde \tau$ describes
the same  observables as $\tau$, with the advantage, however, of coming from
a monodromy matrix made up of ultralocal site operators. Since $B, \tilde B$
and $C, \tilde C$ are respectively different, the diagonalizations of $\tau$
and $\tilde \tau$ by means of Algebraic Bethe Ansatz are made, in principle, on
different (but related) vector spaces. In \cite{FR} we diagonalized $\tau$
and conjectured that it describes, in the continuum limit, the left sector of
minimal CFT's on a cylinder; in a forthcoming publication \cite {35} we will
write the Bethe equations and the eigenvalues/eigenvectors of the transfer
matrix $\tilde \tau$ and we will show that they coincide with the homonymous
quantities in the Liouville model \cite {FT}. We will also disentangle the
comparison between the eigenvectors of $\tau$ and $\tilde \tau$.

\medskip

We are going to repeat the {\it unbraiding} procedure making use of the right
Lax operators $\bar L_m(\la)$ (\ref {lax}). At first, we write $\bar L_m(\la)$
in terms of ultralocal operators. Using (\ref {Wreal}) we have that
\be
\bar L_{m}(\lambda) =
\left ( \begin{array}{cc} e^{i(x_m-x_{m+1}+p_m)} & \Delta \lambda
e^{i(x_m-x_{m+1}-p_m)}  \\ \Delta \lambda  e^{-i(x_m-x_{m+1}-p_m)} &
e^{-i(x_m-x_{m+1}+p_m)}\\ \end{array} \right ) = D_{m+1}^{-1}\bar U_m(\la)
\, ,
\label {rdecomp}
\ee
where $D_m$ is still given by (\ref {DUdef}) and
\be
\bar U_m(\la)\equiv\left ( \begin{array}{cc} e^{i(x_m+p_m)} & \Delta \lambda
e^{i(x_m-p_m)}  \\ \Delta \lambda  e^{-i(x_m-p_m)} &  e^{-i(x_m+p_m)}\\
\end{array} \right )  \, .
\label{barUdef}
\ee
It is evident that $\bar U_{m}(\la)$ depends only on ultralocal site
variables and hence that formula (\ref {rdecomp}) decomposes the right Lax
operator in its ultralocal components.     Decomposition (\ref {rdecomp})
allows us to rewrite the right monodromy matrix (\ref {rightmon}) as follows:
\ba  \bar M(\la)&=&\bar L_N(\la ^{-1})\bar L_{N-1}(\la ^{-1})...\bar
L_2(\la^{-1})\bar L_1(\la^{-1})=\nonumber \\  &=&[D_1^{-1}\bar
U_N(\la^{-1})]\, [D_N^{-1}\bar U_{N-1}(\la^{-1})]...[D_3^{-1}\bar
U_2(\la^{-1})]\, [D_2^{-1}\bar U_1(\la^{-1})] = \nonumber \\  &=&D_1^{-1}\,
[\bar U_N(\la^{-1})D_N^{-1}]\, [\bar U_{N-1}(\la^{-1})D_{N-1}^{-1}]...[\bar
U_2(\la^{-1})D_2^{-1}]\, [\bar U_1(\la^{-1})D_1^{-1}] \,  D_1 \, . \nonumber
\ea  This means that the matrix  \be   \tilde {\bar M}(\la)\equiv D_1 \bar
M(\la) D_1 ^{-1} \label{rMtil}  \ee  can be written as   \be  \tilde {\bar
M}(\la)=\tilde {\bar L}_N(\la^{-1})\tilde {\bar L}_{N-1}(\la^{-1})...\tilde
{\bar L}_2(\la^{-1}) \tilde {\bar L}_1(\la^{-1}) \, , \label {rtildeM}  \ee
in terms of the ultralocal operators
\be
\tilde {\bar L}_m(\la)\equiv \bar
U_m(\la) D_m^{-1}   =q^{1/4} \left ( \begin{array}{cc} e^{ip_m} & \Delta
\lambda  e^{i(2x_m-p_m)}  \\ \Delta \lambda  e^{-i(2x_m-p_m)} &  e^{-ip_m}\\
\end{array} \right ) \,  .
\label{rLreal}
\ee
From this realization of the operators $\tilde {\bar L}_m(\la)$ and from commutation relations
(\ref {heis}) the Yang-Baxter relation follows:
\be
R_{ab}\left (\frac {\lambda}{\lambda ^\prime}\right)\tilde {\bar L}_{am}(\frac {1}{\lambda }) \tilde {\bar L}_{bm}(\frac {1}{\lambda ^\prime})=
\tilde {\bar L}_{bm}(\frac {1}{\lambda ^\prime})\tilde {\bar L}_{am}(\frac {1}{\lambda})R_{ab}\left(\frac
{\lambda}{\lambda ^\prime} \right) \, . \label{ryb}
\ee
Since $\tilde {\bar L}_m$ are ultralocal, the Yang-Baxter relation
is also true for the matrix $\tilde {\bar M}$,
\be
R_{ab}\left (\frac {\lambda}{\lambda ^\prime}\right)\tilde {\bar M}_a(\lambda)
\tilde {\bar M}_b(\lambda ^\prime)=
\tilde {\bar M}_b(\lambda ^\prime)\tilde {\bar M}_a(\lambda)R_{ab}\left(\frac
{\lambda}{\lambda ^\prime} \right) \, , \label{rMyb}
\ee
and gives rise to the commutativity of the transfer matrix
$\tilde{\bar \tau} (\la)\equiv {\mbox {Tr}} \tilde {\bar M}(\la)$ with itself
for different values of the spectral parameter:
\be
[\tilde {\bar \tau} (\la) , \tilde {\bar \tau} (\la ^\prime) ]=0 \, .
\ee

\medskip

The unbraided monodromy matrix $\tilde {\bar M}$ is directly connected to the
right conformal  monodromy matrix $\bar M$ by (\ref {rMtil}). More
explicitly, the matrix elements of $\bar M$ and $\tilde {\bar M}$,
\begin{equation}
{\bar M}(\lambda )\equiv
\left ( \begin{array}{cc}  {\bar A}(\lambda )& {\bar B}(\lambda)
\\ {\bar C}(\lambda)  & {\bar D}(\lambda) \\ \end{array} \right ) \quad , \quad
\tilde {\bar M}(\lambda )\equiv
\left ( \begin{array}{cc}  \tilde {\bar A}(\lambda )& \tilde {\bar B}(\lambda)
\\ \tilde {\bar C}(\lambda)  & \tilde {\bar D}(\lambda) \\ \end{array} \right ) \,  ,
\ee
are related in this way:
\ba
\bar A(\la)=e^{-ix_1}\tilde {\bar A}(\la)e^{ix_1} \quad &,& \quad \bar B(\la)=e^{-ix_1}
\tilde {\bar B}(\la) e^{-ix_1} \, , \nonumber \\
\bar C(\la)=e^{ix_1}\tilde {\bar C}(\la) e^{ix_1} \quad &,& \quad \bar D(\la)=e^{ix_1}\tilde {\bar D}(\la)e^{-ix_1} \, . \label{rABCD}
\ea
Using the same technique as before -- equations (\ref {scorp}, \ref {rel1}) --
we simplify (\ref {rABCD}) into
\ba
&\bar A(\la)=q^{-1/2}\tilde {\bar
A}(\la) \quad , \quad   \bar D(\la)=q^{-1/2}\tilde {\bar D}(\la)\,   , &
\label {rAD} \\  &\bar B(\la)=q^{-1/2}e^{-2ix_1}\tilde {\bar B}(\la) \quad ,
\quad  \bar C(\la)=q^{-1/2}e^{2ix_1}\tilde {\bar C}(\la)\, .&   \label {rBC}
\ea
In conclusion, we have that
\be
\tilde {\bar \tau }(\la)=\tilde {\bar A}(\la)+\tilde {\bar
D}(\la)=q^{1/2}[\bar A(\la)+\bar D(\la)] = q^{1/2}\bar \tau (\la) \, ,
\ee
{\it i.e.} the transfer matrix $\tilde{\bar \tau} (\la)={\mbox {Tr}} \tilde {\bar M}(\la)$ is
proportional to the transfer matrix ${\bar \tau} (\la)\equiv {\mbox {Tr}} {\bar
M}(\la)$ and  hence describes the same observables. We can comment these
results concerning the right sector in the same way as we have done for those
about the left one, simply turning the word "left" into "right".

\section{From the braided to the usual Yang-Baxter relation: off-critical case}
\setcounter{equation}{0}
We now want to extend our unbraiding procedure by connecting the off-critical
monodromy matrix ${\bf M}$  (\ref {Rightmon}) with a suitable monodromy matrix
satisfying the Yang-Baxter relation. Following what we have just done in the
conformal case, we want to write (\ref {Rightmon}) in terms of ultralocal
operators. Decomposition (\ref{decomp})  for $L_m$ and decomposition
(\ref{rdecomp}) for $\bar L_m$ give:  \ba  {\bf M}(\la)&=& [D_1^{-1}\bar U_N
({\mu ^{\frac {1}{2}}}{\la}^{-1})]\,   [D_N^{-1} \bar U_{N-1}({\mu ^{\frac
{1}{2}}}{\la}^{-1})] \, [D_{N-1}   U_{N-2}(\mu ^{\frac {1}{2}}\la )] \,
[D_{N-2} U_{N-3}(\mu ^{\frac {1}{2}}\la )] \ldots \nonumber \\  &&\ldots
[D_5^{-1}\bar U_4(\mu ^{\frac {1}{2}}{\la}^{-1})]\, [D_4^{-1} \bar U_{3}(\mu
^{\frac {1}{2}}{\la}^{-1})] \, [D_{3} U_{2}(\mu ^{\frac {1}{2}}\la )] \,
[D_{2} U_{1}(\mu ^{\frac {1}{2}}\la )] = \label {unbr} \\  &=& D_1^{-1}[\bar
U_N ({\mu ^{\frac {1}{2}}}{\la}^{-1})   D_N^{-1}]\, [ \bar U_{N-1}({\mu ^{\frac
{1}{2}}}{\la}^{-1}) D_{N-1}] \,    [U_{N-2}(\mu ^{\frac {1}{2}}\la )
D_{N-2}]\, [ U_{N-3}(\mu ^{\frac {1}{2}}\la )D_{N-3}^{-1}] \ldots \nonumber \\
 &&\ldots [\bar U_4(\mu ^{\frac {1}{2}}{\la}^{-1}) D_4^{-1}]\, [ \bar
U_{3}(\mu ^{\frac {1}{2}}{\la}^{-1}) D_{3}]\, [ U_{2}(\mu ^{\frac {1}{2}}\la )
  D_{2}]\, [ U_{1}(\mu ^{\frac {1}{2}}\la ) D_1^{-1}] D_1 \, .  \nonumber  \ea
This means that the matrix
\be
\tilde {\bf M}(\la)\equiv D_1\, {\bf M}(\la)\,  D_1^{-1} \label{bfMtil}
\ee
can be written as
\be
\tilde {\bf M}(\la)=\prod \limits _{i=1}^{\stackrel {N/4}{\leftarrow}} \tilde {\bar L}_{4i}(\mu ^{\frac {1}{2}}{\la}^{-1})\tilde {\bar L}_{4i-1}^\prime (\mu ^{\frac {1}{2}}{\la}^{-1})\tilde {L}_{4i-2}(\mu ^{\frac {1}{2}}{\la})\tilde {L}_{4i-3}^\prime (\mu ^{\frac {1}{2}}{\la}) \, ,
\label {tildebfM}
\ee
in terms of the ultralocal operators (\ref{Ltil1}, {\ref{rLreal}) and
\ba
\tilde {\bar L}_{4i-1}^\prime(\la)&\equiv &\bar U_{4i-1}(\la) D_{4i-1} \, ,  \label{tildeL2} \\
\tilde L_{4i-3}^\prime(\la)&\equiv &U_{4i-3}(\la) D_{4i-3}^{-1} \, . \label{tildeL4}
\ea
The new operators (\ref{tildeL2}) and (\ref{tildeL4}) inherit the
realization
\be
\tilde {\bar L}_m^\prime(\la)=q^{-1/4} \left ( \begin{array}{cc} e^{i(2x_m+p_m)} &
\Delta \lambda  e^{-ip_m}
\\ \Delta \lambda  e^{ip_m} &  e^{-i(2x_m+p_m)}\\ \end{array} \right ) \, ,
\label {prLreal}
\ee
\be
\tilde {L}_m^\prime(\la)=q^{1/4} \left ( \begin{array}{cc} e^{-i(2x_m-p_m)} & \Delta \lambda  e^{-ip_m}
\\ \Delta \lambda  e^{ip_m} &  e^{i(2x_m-p_m)}\\ \end{array} \right ) \, .
\label {pLreal}
\ee
We already know -- formul{\ae} (\ref {yb}, \ref {ryb}) -- that operators
$\tilde L_{m}(\la)$ and $\tilde {\bar L}_{m}(\la ^{-1})$ satisfy the Yang-Baxter relation and
this is also true for operators $\tilde L_{m}^\prime(\la)$ and $\tilde {\bar
L}_{m}^\prime(\la^{-1})$, from direct calculation which uses (\ref
{prLreal}) and (\ref {pLreal}). Since all these operators are ultralocal, the
unbraided matrix $\tilde {\bf M}(\la)$ satisfies also the Yang-Baxter
relation,  \be
R_{ab}\left (\frac {\lambda}{\lambda ^\prime}\right)\tilde {\bf
M}_a(\lambda)  \tilde {\bf M}_b(\lambda ^\prime)=  \tilde {\bf M}_b(\lambda
^\prime)\tilde {\bf M}_a(\lambda)R_{ab}\left(\frac  {\lambda}{\lambda ^\prime}
\right) \, ,  \label{bfMyb}
\ee
and produces a transfer matrix $\tilde {\bf t} (\la)\equiv {\mbox {Tr}}\, \,
\tilde {\bf M}(\la)$  commuting with itself for different values of the
spectral parameter:   \be  [\tilde {\bf t} (\la) , \tilde {\bf t} (\la
^\prime) ]=0 \, .  \ee

\medskip

Now, we analyse in more detail the relation between the off-critical monodromy
matrix  ${\bf M}$ and the monodromy matrix $\tilde {\bf M}$. From formula
(\ref {bfMtil}) it follows that this relation formally coincides with relation (\ref {rMtil})
between $\bar M$ and $\tilde {\bar M}$. Therefore, we need to rewrite properly (\ref {rABCD})
in the form:
\ba
&{\bf A}(\la)=e^{-ix_1}\tilde {\bf A}(\la)e^{ix_1}  \quad , \quad
{\bf B}(\la)=e^{-ix_1}\tilde {\bf B}(\la) e^{-ix_1}\, , & \label {bfAB} \\
&{\bf C}(\la)=e^{ix_1}\tilde {\bf C}(\la)e^{ix_1} \quad , \quad
{\bf D}(\la)=e^{ix_1}\tilde {\bf D}(\la)e^{-ix_1}\, . &  \label {bfDC}
\ea
Besides, as in that case we can exchange the factors $e^{\pm i x_1}$ with the
elements $\tilde {\bf A}$, $\tilde {\bf B}$, $\tilde {\bf C}$, $\tilde {\bf
D}$ of $\tilde {\bf M}$:
\ba
&{\bf A}(\la)=q^{-1/2}\tilde {\bf A}(\la) \quad , \quad   {\bf
D}(\la)=q^{-1/2}\tilde {\bf D}(\la)\, , &  \label {bfAD1} \\  &{\bf
B}(\la)=q^{-1/2}e^{-2ix_1}\tilde {\bf B}(\la) \quad , \quad   {\bf
C}(\la)=q^{-1/2}e^{2ix_1}\tilde {\bf C}(\la)\, .&  \label {bfBC1}  \ea
As in the conformal case, we remark that ${\bf t}(\la)\equiv {\mbox {Tr}}\, \, {\bf M}(\la)$ and $\tilde {\bf t}(\la)= {\mbox {Tr}}\, \, \tilde {\bf M}(\la) $ are proportional:
\be
\tilde {\bf t} (\la)=q^{1/2} {\bf t}(\la) \, .
\ee
Hence they describe the same observables. Likewise, since $\bf B, \tilde {\bf
B}$ and $\bf C, \tilde {\bf C}$ are different, the diagonalization of $\bf t$
and $\tilde {\bf t}$ by means of Bethe Ansatz is made, in principle, on
different (but related) vector spaces. In \cite {FR} we diagonalized $\bf t$
and conjectured that it describes, in the cylinder continuum limit, minimal
CFT's perturbed by the $\Phi_{1,3}$ primary operator.  In a forthcoming
publication \cite {35} we will write the Bethe equations and the
eigenvalues/eigenvectors of the transfer matrix $\tilde {\bf t}$ and we will
show that they coincide with the Bethe equations and the
eigenvalues/eigenvectors of the transfer matrix for the Lattice sine-Gordon
model \cite {IK}. Since minimal CFT's perturbed by $\Phi_{1,3}$ are a
restriction of the sine-Gordon model \cite {LCL}, the Bethe Ansatz
construction of the eigenvectors of $\bf t$ and $\tilde {\bf t}$ will be
useful to prove our conjecture.

\medskip

Eventually, we give briefly analogous results for the other off-critical
monodromy matrix (\ref {Leftmon}) with entries defined by:
\begin{equation}
{\bf M}^\prime (\lambda )\equiv  \left ( \begin{array}{cc}  {\bf
A}^\prime(\lambda )& {\bf B}^\prime (\lambda)  \\ {\bf C}^\prime (\lambda)  &
{\bf D}^\prime (\lambda) \\ \end{array} \right ) \, .
\end{equation}
Through the same unbraiding procedure we end up with a
monodromy matrix built up by ultralocal site operators
\be
\tilde {\bf M}^\prime (\la)\equiv D_1^{-1}{\bf
M}^\prime (\la) D_1 \, ,
\label{bfMtil'}
\ee
which can be written as follows:
\be  \tilde {\bf M}^\prime(\la)=\prod \limits _{i=1}^{\stackrel
{N/4}{\leftarrow}} \tilde {L}_{4i}(\mu ^{\frac {1}{2}}{\la})\tilde
{L}_{4i-1}^\prime (\mu ^{\frac {1}{2}}{\la})\tilde {\bar L}_{4i-2}(\mu ^{\frac
{1}{2}}{\la}^{-1})\tilde {\bar L}_{4i-3}^\prime (\mu ^{\frac
{1}{2}}{\la}^{-1}) \, .  \label {tildebfM'}
\ee
Likewise, the matrix $\tilde {\bf M}^\prime (\la)$ satisfies the Yang-Baxter
relation and the connection between its entries and those of the monodromy
matrix ${\bf M}^\prime (\la)$ writes down explicitly
\ba
&{\bf A}^\prime(\la)=q^{1/2}\tilde {\bf
A}^\prime (\la) \quad , \quad  {\bf D}^\prime (\la)=q^{1/2}\tilde {\bf
D}^\prime (\la)\, , & \label {bfAD1'} \\  &{\bf B}^\prime
(\la)=q^{1/2}e^{2ix_1}\tilde {\bf B}^\prime (\la) \quad , \quad   {\bf
C}^\prime (\la)=q^{1/2}e^{-2ix_1}\tilde {\bf C}^\prime (\la)\, .&  \label
{bfBC1'}
\ea
Therefore, ${\bf t}^\prime (\la)\equiv {\mbox {Tr}}\, \, {\bf
M}^\prime (\la)$ and $\tilde {\bf t}^\prime (\la)\equiv {\mbox {Tr}}\, \,
\tilde {\bf M}^\prime (\la) $  are proportional,  \be  \tilde {\bf t}^\prime
(\la)=q^{-1/2} {\bf t}^\prime (\la) \, ,  \ee  and consequently describe the
same observables.    In a forthcoming publication \cite {35} we will write the
Bethe equations and the eigenvalues/eigenvectors of $\bf {\tilde t}^\prime $
and we will show that they coincide with the homonymous quantities for $\bf
{\tilde t}$ and hence  with the homonymous quantities for the lattice
sine-Gordon model \cite{IK}.

\medskip

{\bf Observation.}
It is important to remark that the existence the of ultralocal Lax operators
(\ref {Lreal}, \ref {rLreal}), satisfying the Yang-Baxter relations (\ref
{yb}, \ref {ryb}), allows us to define, if $N$ is even, the {\it factorised}
monodromy matrix
\ba
{\bf M}^F(\la)&=&\tilde {L}_{N}(\mu ^{1/2}\lambda )\tilde
{L}_{N-2}(\mu ^{1/2}\lambda )\ldots \tilde {L}_{4}(\mu ^{1/2}\lambda )\tilde
{L}_{2}(\mu ^{1/2}\lambda )\,  \cdot \nonumber \\  &&\cdot  \, \, \tilde {\bar
L}_{N-1}(\mu ^{1/2}\la ^{-1}) \tilde {\bar L}_{N-3}(\mu ^{1/2}\la^{-1})
\ldots  \tilde {\bar L}_{3}(\mu ^{1/2}\la^{-1}) \tilde {\bar L}_{1} (\mu
^{1/2}\la^{-1}) \label{M} \, ,
\ea
which satisfies the Yang-Baxter relation:
\be
R_{ab}\left (\frac {\lambda}{\lambda ^\prime}\right){\bf
M}_a^F(\lambda)  {\bf M}_b^F(\lambda ^\prime)=  {\bf M}_b^F(\lambda
^\prime){\bf M}_a^F(\lambda)R_{ab}\left(\frac  {\lambda}{\lambda ^\prime}
\right) \, .
\label{MYB}
\ee
Matrix (\ref {M}) is made up of two parts. The left part contains only
operators  $\tilde L_{2m}$ on even sites, the right part only operators
$\tilde {\bar L}_{2m+1}$ on odd sites. From formul{\ae}
(\ref{decomp}, \ref{Ltil1}) and  (\ref {rdecomp}, \ref {rLreal}) it follows
that $\tilde L_{2m}=D^{-1}_{2m+1}L_{2m}D_{2m}$ and that $\tilde {\bar
L}_{2m+1}=D_{2m+2}{\bar L}_{2m+1}D_{2m+1}^{-1}$, where $D_m$, given by
(\ref{DUdef}), is a simple diagonal matrix. Since in the scaling limit left
and right Lax operators $L_{2m}$ and ${\bar L}_{2m+1}$ become completely
chiral and antichiral respectively (see section 9 of \cite {FR}), in the same
limit matrix (\ref{M}) is the product of two matrices, one depending mainly on
the Feigin-Fuks boson $\phi$, the other depending mainly on $\bar \phi$. These
matrices, however, are not completely chiral, because of the presence of the
diagonal matrices $D_m$ inside them. Nevertheless, directly in the scaling
limit ($\Delta\rightarrow 0$) a chiral part by anti--chiral part factorised
monodromy matrix was also discovered in \cite{BLZ} to satisfy the Yang-Baxter
relation. Apparently, that matrix is slightly different from the scaling limit
of (\ref{M}) and from the lattice we did not find, by now, any way to reproduce
exactly that matrix, preserving Yang-Baxter relation. Therefore, in order to
compare our factorised monodromy matrix with that in \cite{BLZ}, we need to
diagonalize by means of Algebraic Bethe Ansatz the transfer matrix ${\bf
t}^F(\la)\equiv{\mbox {Tr}} {\bf M}^F(\la)$, which, as a consequence of (\ref
{MYB}), commutes for different values of the spectral parameter. This will be
the issue of a forthcoming publication \cite{35}.

\section{Perspectives}
We have found a Yang-Baxter relation in theories controlled by a braided
Yang-Baxter algebra, working out the construction in
physical examples involving the lattice (m)KdV theory as main ingredient.
Since this theory is the prototype of non-ultralocal theories and we have gone
from non-ultralocal commutators to ultralocal ones, our treatment is
completely general for what concerns lattice theories (and their continuum
limit). For instance, our method could be applied to the quantum theory
described in \cite{FRS} -- which exhausts all the other integrable
perturbations of minimal CFT's --  or to $W_3$ symmetric
CFT's \cite{ZW3} as described in \cite{BHK}. Moreover, this unbraiding
transformation has been also formulated under a general point of view giving
rise to an algebra isomorphism provided a matrix $D$, satisfying suitable
conditions, exists. In the next future the proposed unbraided monodromy
matrices -- especially the factorised monodromy matrix -- will be worthy of
being investigated. Indeed, we will better understand the spectrum of braided
and unbraided theories by comparing the Algebraic Bethe Ansatz representations
\cite{35}. In this way it will be also possible to perform the (cylinder)
continuum limit and obtain the spectrum of field theories (in finite volume,
{\it i.e.} at finite temperature).

\medskip

{\bf Acknowledgments} - D.F. is grateful to F. Colomo, E. Ercolessi and M.
Stanishkov for discussions and thanks EPRSC for the grant GR/M66370. M.R.
thanks EPRSC for the grant GR/M97497. The work has been partially supported by
EC TMR Contract ERBFMRXCT960012.


\begin{thebibliography}{99}

\bibitem{STF}
E.K. Sklyanin, L.A. Takhtajan, L.D. Faddeev, \TMP{40} (1980) 688;

\bibitem{KBI}
V.E. Korepin, N.M. Bogoliubov, A.Z. Izergin, {\it Quantum
inverse scattering  method and correlation functions}, Cambridge University
Press, Cambridge  (1993), \\
R.J. Baxter, {\it Exactly solved models in
statistical mechanics}, Academic  Press, London (1982);

\bibitem{Dr1}
V.G. Drinfeld, {\it Quantum groups} in Proceedings of the International
Congress of Mathematicians, Berkeley, AMS (1987) 798 and Sov. Math. Dokl.
{\bf32} (1985) 254,  \\
M. Jimbo, \LMP{10} (1985) 63;

\bibitem{FRT}
L.D. Faddeev, N.Yu. Reshetikhin, L.A. Takhtajan, Alg. Anal. {\bf  1} (1989)
178;

\bibitem{H}
L. Hlavaty, J. Math. Phys. {\bf 35} (1994) 2560, \\  L. Hlavaty, A. Kundu,
Int. J. Mod. Phys {\bf A11} (1996) 2143;

\bibitem{K}
A.Kundu, \MPL{A10} (1995) 2955;

\bibitem{FR}
D. Fioravanti, M. Rossi, {\it A braided Yang-Baxter Algebra in
a Theory of two coupled Lattice Quantum KdV: algebraic properties and ABA
representations},  hep-th/0104002;

\bibitem{G}
J.L. Gervais, \PL{B160} (1985) 277, \\
B.A. Kupershmidt, P. Mathieu, \PL{B227} (1989) 245;

\bibitem{BLZ}
V.V. Bazhanov, S.L. Lukyanov, A.B. Zamolodchikov, \CMP{177} (1996) 381, \CMP{190} (1997) 247,
\NP{B489} (1997) 487, \CMP{200} (1999) 297;

\bibitem{FS}
D.Fioravanti, M.Stanishkov, Phys.Lett. {\bf B430} (1998) 109;

\bibitem{BPZ}
A.A. Belavin, A.M. Polyakov, A.B. Zamolodchikov, \NP{B241}
(1984) 333;

\bibitem{AZ}
A.B. Zamolodchikov, Adv. Stud. Pure Math. {\bf 19} (1989) 641;

\bibitem{LCL}
A. LeClair, Phys.Lett. {\bf B230} (1989) 103;

\bibitem{FF}
B.L. Feigin, D.B. Fuks, Lect. Notes in Math., vol {\bf 1060} Springer (1984), \\
V.Dotsenko, V.Fateev, Nucl. Phys. {\bf B240} (1984) 312;

\bibitem{FV}
L. Faddeev, A.Yu. Volkov, Lett. Math. Phys. {\bf 32} (1994) 125, Phys. Lett. {\bf B315} (1993) 313;

\bibitem{35}
D. Fioravanti, M. Rossi, work in progress;

\bibitem{FT}
L.D. Faddeev, O. Tirkkonen, \NP{B453} (1995) 647;

\bibitem{IK}
A. Izergin, V. Korepin, \NP{B205} (1982) 401;

\bibitem{FRS}
D. Fioravanti, F. Ravanini, M. Stanishkov, \PL{B276} (1996) 79;

\bibitem{ZW3}
A.B. Zamolodchikov, \TMP{65} (1985) 1205;

\bibitem{BHK}
V.V. Bazhanov, A.N. Hibberd, S.M. Khoroshkin,
{\it Integrable structure of $W_3$ Conformal Field Theory, Quantum Boussinesq Theory and
Boundary Affine Toda Theory}, hep-th/0105177.

\end{thebibliography}
\end{document}